\documentclass[twocolumn,showpacs,preprintnumbers,amsmath,amssymb]{revtex4}

\usepackage{epsfig}
\usepackage{graphicx}
\usepackage{dcolumn}

\newcommand{\beq}{\begin{equation}}
\newcommand{\eeq}{\end{equation}}
\newcommand{\beqa}{\begin{eqnarray}}
\newcommand{\eeqa}{\end{eqnarray}}

\newcommand{\om}{\omega}

\newcommand{\ra}{\rangle}

\def\jpb#1{{ J.\ Phys.\ B} {\bf#1}}

\def\pra#1{{ Phys.\ Rev. A\/} {\bf#1}}

\def\prl#1{{ Phys.\ Rev.\ Lett.} {\bf#1}}
\def\qic#1{{ Quant.\ Inf.\ Comp.} {\bf#1}}
\def\sci#1{{ Science} {\bf#1}}

\begin{document}

\title{Reverse Engineering with Quantum Noise}

\author{Muhammed Y\"ona\c{c} and Joseph H. Eberly}
\email{yonac@pas.rochester.edu} \affiliation{Rochester Theory Center, and Department of Physics and
Astronomy, University of Rochester, New York 14627, USA }

\begin{abstract}
We show that specific quantum noise, acting as an open-system reservoir for non-locally entangled atoms, can serve to preserve rather than degrade joint coherence. This creates a new type of long-time control over hiding and recovery of quantum entanglement.
\end{abstract}

\maketitle

\section{Introduction}

Usually the ``open" part of a universe is associated with a reservoir or a noise source and is the part that is poorly known or resistant to control. The quantum vacuum is the most fundamental example. Here we examine a situation in which the opposite is the case, in which the open part of the relevant universe, the noise reservoir, is known to be quantum radiation and its interaction is used to control and manage, actually store safely, pre-existing coherence. This is what could be called reverse engineering with quantum noise. A case of particular interest arises when the coherence is shared between parts of the system. For example, it has been shown recently \cite{Yu-EberlyPRL06} that open quantum system theory has unanticipated decoherence features such as non-additive response to weak Markovian noise when the coherence is a two-party entanglement. Here we demonstrate an example of reverse engineering using quantum noise, the consequence of which is a long-term on/off quantum switch for entanglement.

The effect of radiation fields on entanglement of CQED atoms has a long history, beginning with the atom-field entanglement exhibited in the original Jaynes-Cummings (JC) paper \cite{Jaynes-Cummings}. Much later work by Gea-Banacloche \cite{Gea-Banacloche} and Phoenix and Knight \cite{Phoenix-Knight} extended these considerations into the long-time domain that was revealed by the discovery of collapse and revival physics \cite{Eberly-etal80, Narozhny-etal, Yoo-Eberly}. Still later, the evolution of CQED entanglement was further extended to the case where the atom and its field start from mixed states \cite{Rendell-Rajagopal}.

We note that an entirely new set of opportunities arises when one asks about the entanglement of two atoms in different CQED cavities. Then  the atoms cannot interact in any way, and pure non-local entanglement enters the picture.  Examinations of this situation have taken into account the back action of the radiation emitted by the atoms themselves, and have included treatments of the cavities ranging between two extremes: broadband absorptive \cite{Yu-Eberly04, Ikram-etal} or mirror-like \cite{Yonac-etal06}. Among the consequences was so-called entanglement sudden death (ESD -- see \cite{Zyczkowski-etal, Daffer-etal, Diosi, Yu-Eberly04}), which in the mirror-cavity case and its analogs was followed by rebirth, with death and rebirth repeated, usually periodically at the JC Rabi rate \cite{Rebirth}. Extensive overviews are given in \cite{Yu-EberlySci09, Das-Agarwal}.

However, a different approach to the rebirth issue is more rewarding. Suppose one applies a modest-strength quantum field in each cavity, sufficient to dominate the JC sequence of $\pm 1$ photon exchanges with the atoms. As mentioned, the local atom-photon entanglement in each cavity has been examined in this case \cite{Gea-Banacloche,Phoenix-Knight}, but the delocalized atom-atom  entanglement has so far been ignored. It is potentially much more valuable in a variety of quantum information storage and/or transport applications.

\section{Evolution Equations}

Quantum field irradiation with separate coherent states is known to induce very complicated $AB$ two-party dynamics \cite{Yonac-Eberly08,Yonac-Eberly09}. Here we analyze this simplest coherent but quantum case. With a calculational trick we obtain a  formula that suggests the surprising possibility of on/off switching of non-local entanglement over times of operation that are much longer than the Rabi cycle time. A direct consequence will be the ability to ``hide" entanglement for a substantial time and recover it deterministically, and if not used hide it again, repeatedly.

The description of CQED evolution is via the Jaynes-Cummings \cite{Jaynes-Cummings} interaction, which is governed by the familiar Hamiltonian (with $\hbar = 1$):
\beqa
H_{\rm tot} &=& \frac{\om_0}{2}\sigma_z^A  +
g(a^{\dagger}\sigma_{-}^A + \sigma_{+}^A a)  + \om a^{\dagger}a \nonumber\\
 && + \frac{\om_0}{2}\sigma_z^B + g(b^{\dagger}\sigma_{-}^B +
 \sigma_{+}^B b) + \om b^{\dagger}b,
\eeqa
where $\om_0$ is the transition frequency between the two levels of the atoms, $g$ is the constant of coupling between the atoms and the fields and $\om$ is the angular frequency of the single-mode field. The usual Pauli matrices describe the atoms, while $a^{\dagger},\ a$ and $b^{\dagger},\ b$ are the raising and lowering operators for the fields in the two single-mode cavities. A case in which the frequencies and detunings are not equal in the cavities has been examined \cite{asymm}.

The JC eigenstates, as superpositions of the bare atom and cavity product states $|g; n\ra$ and $|e; n-1\ra$ are well known \cite{Jaynes-Cummings}. In order to calculate the time evolution of a joint atom-atom state we need to first calculate the time evolution of the states of the individual sites, and for either site $A$ or site $B$ we have:
\beqa
e^{i H_I t}|e;n\rangle &=& \cos(gt\sqrt{n+1})|e;n\rangle \nonumber \\
&-& i\sin(gt\sqrt{n+1})|g;n+1\rangle \\
e^{i H_I t}|g;n\rangle &=& \cos(gt\sqrt{n})|g;n\rangle \nonumber \\
&-& i\sin(gt\sqrt{n})|e;n-1\rangle.
\eeqa

\section{Two-Qubit Theory with Coherent State Fields}

We assume $AB$ quantum information has been stored (qubit entanglement has been arranged) prior to $t=0$, for example in the pure Bell State
\beq \label{BellState}
|\Psi_{AB}(0)\rangle = (|eg\rangle+|ge\rangle)/\sqrt{2}.
\eeq
The coherent state characterized by $\bar{n}=|\alpha|^2$ is given by
$|\alpha\rangle=\sum_{n=0}^{\infty}A_n |n\rangle$, where $A_n = e^{-|\alpha|^2/2}\alpha^{n}/\sqrt{n!}$. Our initial state for the whole system is therefore
\beq
|\Psi_{tot}(0)\rangle = |\Psi_{AB}(0)\rangle \otimes |\alpha\rangle \otimes |\alpha\rangle.
\eeq

Using these results, the time evolution of the initial state of the system is found to be given by the double sum,
\beq
|\Psi_{tot}(t)\rangle = \frac{1}{\sqrt{2}} \sum_{n=0}^{\infty}\sum_{m=0}^{\infty} A_n A_m \Big(K_{mn}\Big),
\eeq
where $K_{mn}$ is given by the formidable expression
\beqa
K_{mn} &=& -iC_{n+1}S_m|e,e;n,m-1\rangle
+ C_{n+1}S_{m}|e,g;n,m\rangle \nonumber\\
&-&S_{n+1}S_n|g,e;n+1,m\rangle -iS_{n+1}C_n|g,g;n+1,m\rangle \nonumber\\
&-&iS_nC_{m+1}|e,e;n-1,m+1\rangle \nonumber \\
&-&S_nS_{m+1}|e,g;n-1,m+1\rangle\nonumber\\
&+&C_nC_{m+1}|g,e;n,m+1\rangle \nonumber \\
&-&iC_nS_{m+1}|g,g;n,m+1\rangle.
\eeqa
with the abbreviations $C_n=\cos(gt\sqrt{n})$ and $S_n=\sin(gt\sqrt{n})$.

By tracing the photon states from $|\Psi_{tot}(t)\rangle\langle \Psi_{tot}(t)|$ we obtain the $4 \times 4$ reduced density matrix $\rho_{AB}$ for the two atom qubits, whose entanglement we will follow. Because of the infinite range of photon numbers in a coherent state, this density matrix is drastically different from the mostly-zero $X$-matrix \cite{Yu-EberlyQIC07} found in almost all prior discussions of rebirths, becoming a matrix with no zero elements at all. That is we have:
\beq
\label{mixedstate0}
\rho_{AB} = \left[
\begin{array}{cccc}
a & 0 & 0 & w \\
0 & b & z & 0 \\
0 & z^* & c & 0 \\
w^* & 0 & 0 & d
\end{array} \right] \to
\left[
\begin{array}{cccc}
      a & x & x & x \\
      x & b & z & x \\
      x & z* & c & x \\
      x & x & x & d \\
    \end{array} \right]. \\
\eeq
However, by adopting a trick described below, whose validity has to be checked numerically, the elements of $\rho_{AB}$ marked $x$ can all be set to zero. This doesn't eliminate the doubly infinite sums, but it  provides a simplification sufficient to lead to a relatively compact final formula, as follows.

We adopt Wootters' concurrence $C$ \cite{Wootters}  ($1 \ge C \ge 0$), where $C=0$ indicates separability (zero entanglement) and $C=1$ means maximal pure state entanglement, as in a Bell state. The concurrence of an $X$ state like (\ref{mixedstate0}) with $x=0$ everywhere is given by the simple expression \cite{Yu-EberlySci09}
\beq \label{generalC}
C =  2\max\{0, |z| - \sqrt{ad} \}.
\eeq
A different approach to measuring the entanglement, via untraced pure states, gives analogous information (see \cite{Qian-Eberly}).

The elements $z,\ a,\ d$ of $\rho_{AB}$ are given by doubly infinite summations. For $z$ one finds
\beqa \label{element_z}
z &=& \frac{1}{2}\Big\{\sum_{n,m}^{\infty} A_n^2A_m^2C_nC_{n+1}C_mC_{m+1}\nonumber\\
&-&A_nA_{n-1}A_mA_{m+1}S_nC_{n+1}C_mS_{m+1}\nonumber\\
&+&A_nA_{n-2}A_mA_{m+2}S_nS_{n-1}S_{m+1}S_{m+2}\nonumber\\
&-&A_nA_{n-1}A_mA_{m+1}S_nC_{n-1}S_{m+1}C_{m+2}\Big\},
\eeqa
and the series summations for $a$ and $d$ are;
\beqa \label{element_a}
a &=& \frac{1}{2}\Big\{\sum_{n,m}^{\infty} A_n^2A_m^2C_{n+1}^2S_m^2\nonumber\\
&+&A_nA_{n+1}A_mA_{m-1}S_{n+1}C_{n+1}S_mC_m\nonumber\\
&+&A_n^2A_m^2S_n^2C_{m+1}^2\nonumber\\
&+&A_nA_{n-1}A_mA_{m+1}S_nC_nS_{m+1}C_{m+1}\Big\}
\eeqa
and
\beqa \label{element_d}
d &=& \frac{1}{2}\Big\{\sum_{n,m}^{\infty} A_n^2A_m^2S_{n+1}^2C_m^2\nonumber\\
&+&A_nA_{n+1}A_mA_{m-1}S_{n+1}C_{n+1}S_mC_m\nonumber\\
&+&A_n^2A_m^2C_n^2S_{m+1}^2\nonumber\\
&+&A_nA_{n-1}A_mA_{m+1}S_nC_nS_{m+1}C_{m+1}\Big\}. \eeqa
The infinite extent of these summations reflects the fact
that we have coupled the qubits to an open state space.

The sums cannot be evaluated in closed form, but Stirling's formula, $n!=\sqrt{2\pi n} n^n e^{-n}$, and Euler's formula for approximating summations by integrals, can be used for coherent states that are even only moderately strong, i.e., $\alpha \ge 10$. The dominant contribution near $\bar n$ because of the Poisson-peaked nature of $A_n$ also justifies the approximation
\beq
\sqrt{n+1}=\sqrt{n}+\frac{1}{2\sqrt{n}}.
\label{appr_sqrt}
\eeq

With these approximations, and the saddle point method of integration, highly simplified expressions for the sums can be found \cite{YonacThesis}. Abbreviating $\tau = gt$, we find, for example,
\beqa \label{int1}
I(\tau) & = & \int_0^{\infty} e^{-\alpha^2}\frac{\alpha^{2n}}{\sqrt{2\pi n}} \frac{e^n}{n^{n}} e^{i\tau/2 \sqrt{n}} dn \nonumber \\
& \cong & \exp\Big (-\frac{\tau^2}{32\alpha^4}\Big ) e^{i\tau/2\alpha}.
\eeqa
A second integral is similar, but with $\exp(i\tau/2 \sqrt{n})$ replaced by $\exp(2i\sqrt{n}\tau)$, and the saddle point method is again appropriate, although evaluation is more complicated. Helpful cancellations \cite{YonacThesis} can be identified and lead to the following expression for $Q(t) = |z|-\sqrt{ad}$:
\beqa \label{C vs T}
Q(t) &\cong&  \frac{1}{4}\Big[\exp\Big (-\frac{\tau^2}{16\alpha^4}\Big)  -1 + e^{-\tau^2/2}\cos(4\alpha\tau)\Big] \nonumber\\
&&+\sum_{k=1,2,...}\frac{1}{2\pi k}\Big[\exp\Big(-\frac{2(\tau-2\pi k\alpha)^2}{1+\pi^2 k^2}\Big) \nonumber \\
&& \hspace{.25in} \times \cos[4\alpha(\tau-2\pi k\alpha)]
\Big].
\eeqa
In writing this last equation we have used the fact that around $\tau=2\pi k\alpha$ only the term with the corresponding $k$ gives a significant contribution to the sums.  The contribution to $\tau=2\pi k\alpha$ from any other $k^{\prime}$ is proportional to $exp\{-4\pi^2 \alpha^2 (k-k^{\prime})^2/[1+(\pi k^{\prime})^2]\}$, so it decays exponentially with the distance from $k$.

\section{Analysis and Summary}

There is a substantial amount of analysis behind formula (\ref{C vs T}), and its feasibility relies on the trick that permits the step $x \to 0$ in (\ref{mixedstate0}). The trick is not hard to understand. First, the relatively narrow Poisson distribution of photon numbers in a coherent state suggests replacing $|\alpha\rangle \otimes |\alpha\rangle$ by the Fock state $|n\rangle \otimes |n\rangle$, since we expect all main effects to be concentrated in the near neighborhood of the Poisson peak at $n \sim \bar n = |\alpha|^2$. Second, we assumed that the coherent states were close enough in amplitude to use the same parameter $\alpha$ for both.

\begin{figure}[t!]
\centering
 \includegraphics[scale=0.54]{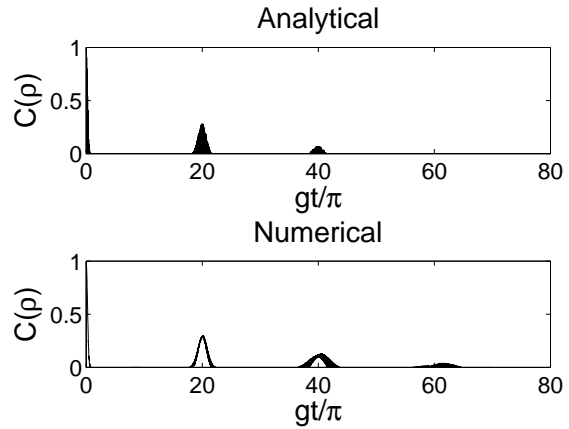}
\caption{\label{num_an} The analytical and numerical results for
entanglement. The former are for the $X$-type $\rho$ while the numerical ones are for the original $\rho$. The shape, location and strength of the revivals are predicted well by analytical formula (\ref{C vs T}).  }
\end{figure}

The combined effect of the local fields induces growth of the elements $a$ and $d$, which are the ones not already present in the density matrix of the original entangled state (\ref{BellState}). Their growth and any decline of $z$ will cause entanglement to decrease. Inspection of formula (\ref{C vs T}) shows that it contains repeated zeros for $Q(t) = |z|-\sqrt{ad}$, meaning repeated deaths and rebirths of entanglement. But between clusters of death and rebirth (\ref{C vs T}) predicts substantial intervals of time when $Q$ remains zero to a very good approximation. This is demonstrated in Fig. \ref{num_an}, where we display concurrence evolution plots (see \cite{Yonac-Eberly09}) from both the trick formula and the corresponding numerical solution to the evolution, without using the trick. Note that concurrence is plotted over a time interval much longer than a Rabi period: $g\Delta\tau \cong \pi$. That is, we have entered the revival time regime explored earlier \cite{Gea-Banacloche, Phoenix-Knight}, but now with results for non-local two-cavity entanglement.

Clearly, both curves in Fig. \ref{num_an} show revivals, i.e.,  repeated eruptions of finite concurrence, separated by relatively long intervals of no appreciable entanglement. This is the specifically quantum effect of the coherent states, acting here as reservoirs in the joint state evolution. As with all quantum revival effects, this is due to the  granularity of the coherent state, to its basis in discrete photon-number states (fractional photons do not exist).  This aspect has been demonstrated experimentally several times (most recently, see \cite{Winelandgroup, Harochegroup}). The two plots are in remarkably good agreement as to position and height of revivals, despite the crude trick played to get the first plot.

The location of the rebirths is controlled by $\alpha$ in the usual way of revivals, and the $kth$ peak height $H_k(\tau)$ decreases with $k$ as given by the formula
\beq\label{rev_env}
2H_k(\tau) \sim 2/{\pi k} - 1 + \exp (-\tau^2/16\alpha^4).
\eeq
In Fig. \ref{num_an,detail}, which shows an expanded snapshot of the curves near to the revival at $\tau = 20\pi$, the analytical formula is seen to predict a micro-structure of repeated ESD events on the Rabi scale. This is not present in the numerically exact plot, which shows no intra-revival death-birth events, and this is true even in the revival immediately following $t=0$.

\begin{figure}[h!]
\includegraphics[scale=0.5]{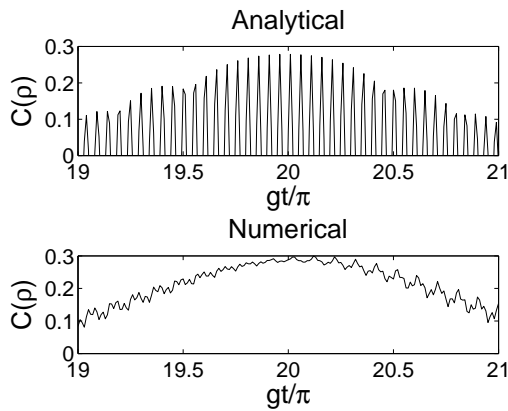}
\caption{ \label{num_an,detail} A detailed plot for $\alpha = 10$ of the results around $t=20\pi/g$.  Analytical results are for the $X$-type $\rho$ while the numerical ones are for the original $\rho$.}
\end{figure}

Figures \ref{num_an} and \ref{num_an,detail} demonstrate what we have claimed, that the quantum nature of the reservoirs, their photonic granularity, can create a new and much longer time scale for management of non-local entanglement. Additionally, on that longer time scale, which is many times the Rabi cycle period, a remarkably complete control of the joint non-local atom-atom state can be achieved. This complete control amounts to an on/off switch for entanglement. Between the revivals entanglement can be considered entirely hidden, but almost fully preserved for recovery at the next revival.

To test that these conclusions are valid generally, and not only for special values of coherent field strength, we show a test in Fig. \ref{revivals 5,6detail} for two smaller values of mean photon number: $\bar n$ = 25 and 36. In those graphs the recovery times are different, and even for those nearer-vacuum fields they still illustrate event-timing control, and one sees the same compact time-zone of non-zero concurrence through almost all of the revival episodes. The modulations on top of the curves can be analyzed, but the key is that they are small and are smaller for higher $\bar n$, as in Fig. \ref{num_an,detail}. In sum, both timing and relative smoothness in C are controllable features.

\begin{figure}[h!]
    \includegraphics[scale=0.4]{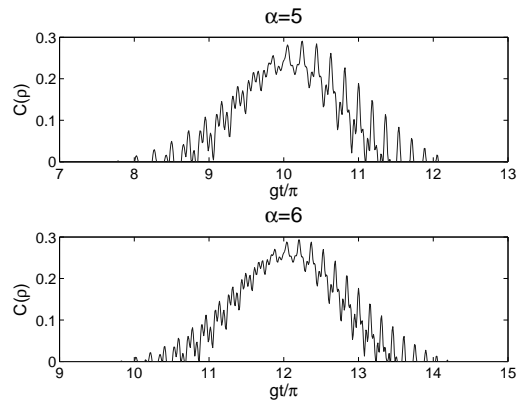}
\caption{Similar to the lower curve in Fig. \ref{num_an,detail} above, except that here $\alpha$ = 5 and 6, rather than 10.}.\label{revivals 5,6detail}
\end{figure}

\newpage

\section{Acknowledgements}

We are pleased to thank Prof. Ting Yu for consultation and collaboration in the early stages of this study. Partial financial support was provided by grants from DARPA  HR0011-09-1-0008 and ARO W911NF-09-1-0385.

\end{document}